\documentstyle[prb,aps,psfig]{revtex}

\newcommand{\ie}{\textit{i.e.}}

\begin{document}
\draft
\title{Size effect in fracture: roughening of crack surfaces and asymptotic analysis}

\author{St\'{e}phane Morel$,^1$  Elisabeth Bouchaud$,^2$ 
and G\'{e}rard Valentin${\,}^1$}

\address{$^1$ Lab. de Rh\'{e}ologie du Bois de Bordeaux, 
        UMR 5103, Domaine de l'Hermitage,
        B.P.10, 33610 Cestas Gazinet, France\\
	$^2$ C.E.A.- Saclay (DSM/DRECAM/SPCSI), 
        91130 Gif-Sur-Yvette Cedex, France}

\maketitle

\begin{abstract}
Recently the scaling laws describing the roughness development of fracture surfaces 
was proposed to be related to the macroscopic elastic energy released during crack 
propagation \cite{Mor00}.  
On this basis, an energy-based asymptotic analysis allows to extend the link to the 
nominal strength of structures.  
We show that a Family-Vicsek scaling leads to the classical size effect of 
linear elastic fracture mechanics. On the contrary, in the case of 
an anomalous scaling, there is a smooth transition from the case of no size effect, 
for small structure sizes, to a power law size effect which appears weaker than 
the linear elastic fracture mechanics one, in the case of large sizes.  
This prediction is confirmed by fracture experiments on wood.
\end{abstract}
%
%\begin{multicols}{2}
%\narrowtext

%
%--Introduction-------------------------------------------------------
%
\protect
\section{Introduction}
\label{sec:intro}
In solid mechanics, an essential scaling problem is the effect of the structure size 
on its nominal strength. This effect is particularly important in the case of
\textit{quasibrittle} materials which are characterized by the existence of 
a large fracture process zone containing many damage microcracks. Materials as different 
as concretes, mortar and rocks, some composites, toughened ceramics and  wood belong to 
this category. Ba\v{z}ant \cite{Baz84,Baz97a} has shown that in this case, contrary to 
what happens for Weibull's statistics \cite{Wei39}, the size effect is linked to the very 
existence of the development of such a large microcracked zone, which implies stress 
redistributions and stored energy release.  

On the other hand, in quasibrittle materials, damage has a strong influence on the local 
deviations of the main crack through its elastic interactions with the microcracks 
\cite{Hor87,Kac94,Law93}. Consequently, the roughness of the fracture surfaces can be 
considered as an inheritance of the damage process and it is naturally tempting to 
correlate their morphology (and especially their fractal properties) to macroscopic 
mechanical properties such as fracture energy or fracture toughness 
\cite{Mos93,Car94,Bou94,Bal96,Bor97,Baz97b,Mor00}.  

In this paper, on the basis of a link recently established \cite{Mor00} between the 
roughening of crack surfaces and the macroscopic energy released during crack propagation 
(this work is summarized in Section \ref{sec:Scaling}), we propose an energy-based 
asymptotic analysis in Section \ref{sec:Sizeffect} which allows to extend this link to the 
nominal strength of large (Sec. \ref{sec:Large_size}) and small (Sec. \ref{sec:Small_size}) 
structures.  An approximate size effect valid everywhere is proposed in 
Sec. \ref{sec:Approx}. This prediction is shown to be in agreement with experimental results 
obtained on wood Sec. \ref{sec:Exp}. Finally, we discuss material-dependent properties in 
Sec. \ref{sec:Conclu}.   
%
%---------------------------------------------------------
%
\protect
\section{Scaling laws of crack surfaces and energy release rate}
\label{sec:Scaling}
The statistical characterization of the fractal morphology of fracture surfaces 
is nowadays a very active field of research.  It is now well established that 
these surfaces, for very different types of materials (from ductile aluminium 
alloys \cite{Dau90,Bou90} to brittle materials like rock 
\cite{Schm94,Plo95,Lop98} or wood \cite{Eng94,Mor98}), exhibit self-affine 
scaling properties in a large range of lengthscales (see \cite{Bou97} for a more 
detailed account of experiments).  Moreover, in addition to this self-affine 
character, recent studies focussed on the complete description (3D) of the morphology 
of crack surfaces \cite{Lop98,Mor98} have shown that the scaling laws governing 
the crack developments in longitudinal and transverse directions are different 
and material dependent \cite{Mor98}.  

Let us consider the development of a fracture surface from a straight 
notch of length $L$ with zero roughness. The mean plane of the crack surface is 
defined as $(x,y)$ where the $x$ axis is perpendicular to the direction of crack 
propagation and the $y$ axis is parallel to this direction.  
For two quasibrittle materials (granite \cite{Lop98} and wood \cite{Mor98}), it has 
been found that the fluctuations $\Delta h$ of the height on the fracture surfaces, 
estimated over a window of size $l$ along the $x$ axis and at a distance $y$ from the 
initial notch exhibited \textit{anomalous} scaling properties which are quite similar to 
those obtained in some models of nonequilibrium kinetic roughening \cite{Das94,Lop97}:
\begin{eqnarray}
\Delta h(l,y) \simeq A \left\{
  \begin{array}{ll}
 	l^{{\zeta}_{loc}} \: \xi (y)^{\zeta -{\zeta}_{loc}}
 	& \mbox{if } \: l \ll \xi (y)\\
 	\xi (y)^{\zeta}
 	& \mbox{if } \: l \gg \xi (y) \label{eq:Anomalous}
  \end{array}
\right.
\end{eqnarray}
where $\xi(y)=By^{1/z}$ depends on the distance $y$ to the initial notch and 
characterizes a crossover length along the $x$ axis. For length scales smaller
than $\xi (y)$, the surface is self-affine, and characterized by the local roughness 
exponent ${\zeta}_{loc}$.  This self-affine character is observed in most experiments 
and a local roughness exponent ${\zeta}_{loc}\simeq 0.8$ is reported in all cases.  
Hence, it has been suggested that this local roughness exponent might be a universal 
index, \ie~independent of the fracture mode and of the material \cite{Bou90}.  

According to Eq. (\ref{eq:Anomalous}), the magnitude of the roughness increases as a 
function of the distance $y$ until the self-affine correlation length $\xi(y)$ reaches 
the system size $L$. This happens at a certain distance $y_{sat}=(L/B)^z$ from the notch: 
$\xi(y\gg y_{sat})=L$. Thus, the first growth regime of the roughness 
(\ie~for $y \ll y_{sat}$) is followed by a stationary regime (for $y \gg y_{sat}$) where 
the magnitude of the roughness remains constant and where the global roughness 
(\ie~measured over the system size $L$) is driven by the global roughness exponent $\zeta$: 
$\Delta h(L,y \gg y_{sat})\sim L^\zeta$.  The main consequence of an anomalous scaling 
[Eq. (\ref{eq:Anomalous})] is that, in this stationary regime, the magnitude of the local 
roughness (\ie~measured on windows $l \ll L$) is not only a function of the window size 
$l$ but also of the system size $L$: 
$\Delta h(l,y \gg y_{sat})\sim l^{{\zeta}_{loc}}\: L^{{\zeta}-{\zeta}_{loc}}$.  

Experimental results obtained on quasibrittle materials have shown the following 
values for the global roughness exponent: $\zeta=1.2$ for granite \cite{Lop98} and 
$1.35$ and $1.60$ for wood (pine and spruce respectively) \cite{Mor98}.  Note that the 
exponent $z$ (called the ``dynamic exponent'') and the prefactors $A$ and $B$ seem also 
to be material-dependent \cite{Mor98}.  

The Family-Vicsek scaling law \cite{Fam91}, where the height fluctuations $\Delta h$ 
scale as:  
\begin{eqnarray}
  \Delta h(l,y)  \simeq  A
        \left\{ \begin{array}{ll}
        l^{{\zeta}_{loc}}   & \mbox{if } \: l \ll \xi (y)\\
        \xi (y)^{{\zeta}_{loc}} & \mbox{if } \: l \gg \xi (y)
\label{eq:Family}
                \end{array}
\right.
\end{eqnarray}
can be seen as a particular case of the anomalous scaling [Eq. (\ref{eq:Anomalous})] where 
$\zeta = \zeta_{loc}$.  In this case [Eq. (\ref{eq:Family})], the magnitudes of the local 
and global roughnesses at saturation, respectively 
$\Delta h(l,y \gg y_{sat})\sim l^{{\zeta}_{loc}}$ and 
$\Delta h(L,y \gg y_{sat})\sim L^{{\zeta}_{loc}}$, are driven by the same roughness exponent 
(the local roughness exponent $\zeta_{loc}$). Furthermore, the local roughness is 
independent of the system size. This is drastically different from what happens in the case 
of an anomalous scaling.  It is worth noticing that both the Family-Vicsek and the anomalous 
scaling laws were used to describe the roughness development of the same granite fracture 
surface \cite{Schm94,Lop98}, and that more accurate results were obtained assuming anomalous 
scaling \cite{Lop98}.    

The anomalous and Family-Vicsek scalings were recently shown to be linked to drastically 
different mechanical behaviors in terms of elastic energy release \cite{Mor00}.  
Within the framework of an equivalent linear elastic problem, a fracture criterion, linking 
the elastic energy release rate $G$ at the macroscale and the fractal nature of the crack at 
the microscale, was proposed: $G\:=\:2\gamma\:\psi(y)/L$.  In this fracture criterion, 
$\gamma$ is the so-called specific surface energy and the ratio $\psi(y)/L$ can be considered 
as a ``roughness factor''. As a matter of fact, it is precisely the ratio of the length of a 
\textit{virtual} crack front over its projected length $L$ (the system size, \ie~the specimen 
width).  This virtual front, parallel to the initial notch, is assumed to be rough only out 
of the average fracture plane $(x,y)$, and this roughness can be defined from the functions 
$\Delta h(l,y)$ at a fixed position $y$ from the initial notch.  

According to an anomalous scaling [Eq. (\ref{eq:Anomalous})], the estimate of the real length 
$\psi$ of the virtual front (which corresponds to the length of a self-affine curve 
\cite{Mor00}) leads to the following expressions of the energy release rate: 
\begin{eqnarray}
        G_R(\Delta a) \:  \simeq  \: 2 \gamma
        \left\{ \begin{array}{ll}
        \sqrt{1+\left( \frac{A B^{\zeta-{\zeta}_{loc}}}
        {{l_o}^{1-{\zeta}_{loc}}} \right)^2 
        {\Delta a}^{2(\zeta-{\zeta}_{loc})/z}} 
        & \mbox{if } \: \Delta a \ll {\Delta a}_{sat}\\
        \sqrt{1+\left( \frac{A}{{l_o}^{1-{\zeta}_{loc}}} \right)^2  
        \:L^{2(\zeta-{\zeta}_{loc})}}
        & \mbox{if } \: \Delta a \gg {\Delta a}_{sat}
\label{eq:Gr}
        \end{array}
\right.      
\end{eqnarray}
where the crack length increments $\Delta a$ and ${\Delta a}_{sat}$ defined from the initial 
notch correspond respectively to the crack positions $y$ and $y_{sat}$ defined in 
Eq. (\ref{eq:Anomalous}).  In the zone where the roughness grows, 
\ie~for $\Delta a \ll {\Delta a}_{sat}$, the fracture equilibrium leads to an energy release 
rate function of the crack length increment $\Delta a$ [Eq. (\ref{eq:Gr})]. The subscript 
$R$ in $G_R$ emphasizes the fact that the resistance to fracture growth is similar to the 
behavior described by a resistance curve (usually called \textit{R}-curve \cite{Law93}). 
Note that the square root terms in Eq. (\ref{eq:Gr}) are dimensionless and correspond to the 
roughness factor $\psi(y)/L$. The term $l_o$ is the lower cutoff of the fractal range of the 
virtual front (\ie~the characteristic size of the smaller microstructural element relevant 
for the fracture process). When the crack increment is large, 
\ie~for $\Delta a \gg {\Delta a}_{sat}$ [Eq. (\ref{eq:Gr})] which corresponds to the 
saturation state of the roughness, the resistance to fracture growth becomes independent of 
the crack length increment because the self-affine correlation length has reached the system 
size: $\xi(\Delta a \gg {\Delta a}_{sat}) \simeq L$.  Introducing the crossover length 
$L_C=({l_o}^{1-{\zeta}_{loc}}/A)^{1/(\zeta-{\zeta}_{loc})}$, the resistance to crack growth 
for large crack length increments becomes:
\begin{equation}
	G_R(\Delta a \gg {\Delta a}_{sat}) \:  \simeq  \: G_{RC} \: 
	\simeq 2 \gamma
	\sqrt{1+\left( \frac{L}{L_C} \right)^{2(\zeta-{\zeta}_{loc})}}  
\label{eq:Grc}        
\end{equation}
where the subscript $C$ in $G_{RC}$ emphasizes that the resistance to crack growth has 
reached an asymptotic or \textit{critical} value.  The main consequence of the link between 
fracture mechanics and anomalous roughening of fracture surfaces \cite{Mor00} is the size 
effect on the critical resistance to crack growth [Eq. (\ref{eq:Grc})]. This size effect is 
the result of the dependence of the maximum magnitude of the roughness at saturation on 
the structure size $L$.  As shown in Fig. \ref{fig:fig1}, the size effect affecting the 
critical resistance $G_{RC}$  exhibits two asymptotic behaviors separated by the crossover 
length $L_C$.  For small structure sizes (\ie~$L \ll L_C$) where the roughness of the fracture 
surface is weak, there is no size effect, and $G_{RC} \simeq 2\gamma$. For large structure 
sizes (\ie~$L \gg L_C$), corresponding to an important fracture roughness, the critical 
resistance evolves as a power law : $G_{RC} \sim L^{\zeta-\zeta_{loc}}$.

In the case of the Family-Vicsek scaling [Eq. (\ref{eq:Family})], on the contrary, the link 
between roughnening of fracture surfaces and material fracture properties \cite{Mor00} 
reduces to a resistance to crack growth independent of the crack increment $\Delta a$ and of 
the specimen size $L$:
\begin{equation}
        G_C (\Delta a) \simeq \: 2 \: \gamma \: 
        \sqrt{1+ \left( \frac{A}{{l_o}^{1-{\zeta}_{loc}}} \right)^2 } 
\label{eq:GcFamily}     
\end{equation}

Thus, an anomalous scaling accounts for an \textit{R}-curve behavior and a size effect on the 
critical resistance to crack growth while a Family-Vicsek scaling reflects a purely elastic 
brittle fracture behavior.   
%
%--Size effect-------------------------------------------------------
%
\protect
\section{Size effect on the nominal strength}
\label{sec:Sizeffect}

We now extend the connection just summarized \cite{Mor00} to the nominal strength of 
structures.  

Within the framework of Ba\v{z}ant's theory \cite{Baz97c}, the size effect (for 
two-dimensional problem) can be described from geometrically similar structures of different 
sizes (with geometrically similar initial cracks or notches) by introducing a nominal stress: 
\begin{equation}
        \sigma_N = \frac{P}{d\:L}
\label{eq:Sn}  
\end{equation}
where $P$ is the external load applied to the structure (considered to be a load independent 
of the displacement), $L$ is the characteristic size of the structure, and $d$ is any length 
of the structure (for instance, as shown in Fig. \ref{fig:fig2} in the case of TDCB specimens, 
$d$ corresponds to the ligament length). When $P=P_u$ which corresponds to the ultimate or 
maximum load, $\sigma_N$ is called the \textit{nominal strength} of the structure.  

On the other hand, at the maximum load $P_u$, the elastic energy release rate $G$ (obtained 
at a constant load $P_u$ or $\sigma_N$) must be equal, according to the linear elastic fracture 
mechanics (LEFM), to the critical resistance to crack growth $G_{RC}$:
\begin{equation}
	G = \frac{1}{L}\biggl [\frac{\partial W^{\ast}}{\partial a}\biggr]_{\sigma_N}
	=\:G_{RC}
\label{eq:G}
\end{equation} 
where the complementary energy $W^{\ast}$ characterizes the energy stored in the structure.  
This energy $W^{\ast}$ can be written in the following form: 
$W^{\ast} = {{\sigma_N}^2}L\:{d\:}^2 f(\alpha)/E$ where $f$ is a dimensionless function 
characterizing the geometry of the structure and $\alpha=a/d$ is the relative crack length.  
Thus, when the maximum load is reached, the nominal strength of the structure can also be 
written as: 
\begin{equation}
	\sigma_N = \sqrt{\frac{E\:G_{RC}}{d\:g(\alpha)}}
\label{eq:Sn-d}
\end{equation} 
where $g(\alpha)=\partial f(\alpha)/\partial \alpha$ corresponds to the dimensionless energy 
release rate function.    

Usually, the main problem with Eq. (\ref{eq:Sn-d}) is to determine the relative crack length 
$\alpha=a/d$ for which the maximum load $P_u$ is reached. Let us consider exclusively a 
structure of ``positive geometry'', \ie~$\partial g(\alpha) /\partial \alpha > 0$  under load 
control. 

If the material exhibits a purely elastic brittle behavior such as the one obtained for a 
Family-Vicsek scaling [Eq. (\ref{eq:GcFamily})], the relative crack length increment $\alpha$ 
at the maximum load corresponds to $\alpha_o=a_o/d$ ($a_o$ being the length of the notch or 
initial crack).  Thus, in the case of a Family-Vicsek scaling, replacing the resistance to 
crack growth $G_C$ by its expression in [Eq. \ref{eq:GcFamily}] and $\alpha$ by $\alpha_o$ 
in Eq. (\ref{eq:Sn-d}), we obtain the following size effect relation: 
\begin{equation}
       \sigma_N \: = \: 
       \sqrt{\frac{2 \gamma E}{m g(\alpha_o)} \: 
       \left[1+ \left( \frac{A}{{l_o}^{1-{\zeta}_{loc}}} \right)^2 \right]^{1/2}}
       \quad L^{-1/2} 
       \label{eq:SizeFam}        
\end{equation}
where $m$ is a proportionality coefficient between $d$ and $L$ ($d=mL$). 
In Eq.(\ref{eq:SizeFam}), the term under the square root is a constant, which means that the 
nominal strength of structures evolves as $\sigma_N \sim L^{-1/2}$. This is in agreement with 
the expected LEFM size effect, \ie~the size effect of a purely elastic brittle fracture 
behavior.  

In the case of an anomalous scaling, the problem appears more complicated.  Indeed, under the 
conditions evoked in the previous case (\ie, structures of positive geometry and under load 
control), if the material exhibits an \textit{R}-curve behavior [Eq. (\ref{eq:Gr})], the crack 
length increment $\Delta a_{sat}$ (limit of the \textit{R}-curve) is actually the limit of 
stability. Thus, the relative crack length at the maximum load can be defined as 
$\alpha=\alpha_o+\theta $ where $\theta=\Delta a_{sat}/d$.  Hence, in the case of an anomalous 
scaling, which reflects the fracture behavior of quasibrittle materials, the knowledge of the 
evolution of the crack length increment $\Delta a_{sat}$ as a function of the structure size is 
the key of the size effect problem.  However, the dependence between $\Delta a_{sat}$ and the 
specimen size does not appear clearly from the roughness analysis \cite{Lop98,Mor98}.  

In order to resolve this problem - which is the central point of this paper -, a possible way 
suggested by Ba\v{z}ant \cite{Baz97c} consists in considering that the failure of a 
quasibrittle material is not only characterized by the specific surface energy $2\gamma$ 
(related to the actual crack surface), but also by a critical damage energy release rate $G_d$ 
per unit volume of \textit{damaged} material (\ie~per unit volume of fracture process zone).  
Thus, one can assume that failure at the maximum load is obtained for the energy balance: 
\begin{equation}
	G_d \: V_{FPZ}= \: 2 \gamma \: A_{r}(\Delta a_{sat})
\label{eq:Crit}
\end{equation} 
where the volume of the fracture process zone can be estimated as 
$V_{FPZ}=L\:{\Delta a_{sat}}^2/n$ with $L\Delta a_{sat}$  the projected crack surface and 
$\Delta a_{sat}\:/n$ the \textit{height} of the process zone (where $n$ is assumed constant, 
\ie~independent of the size $L$).  The surface $A_{r}(\Delta a_{sat})$ corresponds to the real 
fracture surface produced during the crack advance $\Delta a_{sat}$. Note that the fracture 
criterion [Eq. (\ref{eq:Crit})] actually corresponds to an equivalent linear elastic problem 
\cite{Baz97c,Baz90} where the effective size of the fracture process zone at the maximum load 
is assumed equal to the crack length increment for which the resistance to crack growth does 
not follow the \textit{R}-curve [Eq. (\ref{eq:Gr})] but remains constant and equal to the 
critical resistance: $G_R(\Delta a_{sat})=G_{RC}$ [Eq. (\ref{eq:Grc})].  

In the following, on the basis of the fracture criterion defined in Eq. (\ref{eq:Crit}), the 
asymptotic values of $\Delta a_{sat}$ are estimated for large and small structures in order to 
obtain the nominal strength respectively for large and small-size asymptotic expansions.
%
%----Large-size asymptotic expansion--------------------------------------------------------
%
\protect
\subsection{Large-size asymptotic expansion of the size effect}
\label{sec:Large_size}

As previously mentioned, the square root terms in Eq. (\ref{eq:Gr}) and Eq. (\ref{eq:Grc}) 
correspond to the ratio of the virtual crack front length $\psi(\Delta a)$ over its projected 
length $L$.  Moreover, from Eq. (\ref{eq:Grc}), relative to the roughness saturation regime, 
one can obtain the maximum length of the virtual crack front since 
$\psi(\Delta a \gg \Delta a_{sat}) \simeq \psi_{max} \simeq  
L \sqrt{1+(L/L_C)^{2(\zeta-{\zeta}_{loc})}}$ \cite{Mor00}.  Thus, for large structure sizes, 
\ie~$L\gg L_C$, the asymptotic value of the real crack surface $A_r$ produced for a crack 
advance $\Delta a_{sat}$ can be estimated from $\psi_{max}$ as: 
$A_r \simeq \beta \: L \: \Delta a_{sat} \: (L/L_C)^{\zeta-\zeta_{loc}}$ where $\beta$ is a 
constant (function of the scaling exponents).  Substituting $A_r$ in Eq. (\ref{eq:Crit}) 
yields the expression of the crack length increment:
\begin{equation}
	\Delta a_{sat} = c^{\ast} \beta \:
	\biggl ( \frac{L}{L_C}\biggr)^{\zeta-\zeta_{loc}}
\label{eq:dasat-l}
\end{equation} 
where $c^{\ast}=n 2\gamma/G_d$ is a material-dependent length scale.  Thus, for large structure 
sizes, the relative size of the fracture process zone (\ie~$\theta=\Delta a_{sat}/d$) is 
expected to evolve as a power law $\theta \sim L^{\zeta-\zeta_{loc}-1}$.  Moreover, from the 
values of the scaling exponents obtained in the roughness analysis of quasibrittle materials 
\cite{Lop98,Mor98}, it appears that the relative size of the process zone becomes negligible 
when the system size increases: $lim \: \theta =0$ for $L\rightarrow +\infty$.  In other terms, 
in large structures, the process zone is expected to lie within only an infinitesimal volume 
fraction of the body and so $lim \: \alpha = \alpha_o$ for $L\rightarrow +\infty$. Note that 
this result is in agreement with Ba\v{z}ant's assumption \cite{Baz97c}.  Hence, the 
dimensionless energy release rate function $g(\alpha)$ being generally a smooth function, 
we may expand it into a Taylor series around $\alpha=\alpha_o$ and Eq.(\ref{eq:Sn-d}) thus 
yields :
\begin{eqnarray}
       \sigma_N & = & \sqrt{\frac{E\:G_{RC}}{d}} \:
       \biggl [ \: g(\alpha_o) + g_1(\alpha_o) \: \theta + 
       g_2(\alpha_o) \: \frac{\theta^2}{2!} +
       g_3(\alpha_o) \: \frac{\theta^3}{3!} + ...\: \biggr ]^{-1/2} \label{eq:LSexp1}\\
       		& = & \sigma_{M} \: 
       \sqrt{\frac{\biggl (1 \: + \: 
       \bigl (\frac{L}{L_C} \bigr)^{2(\zeta-\zeta_{loc})} \biggr )^{1/2}}
       {\frac{L}{L_1} \: + \: {\bigl (\frac{L}{L_C}\bigr )}^{\zeta-\zeta_{loc}} \: 
       + \: b_2 \frac{L_1}{L} {\bigl (\frac{L}{L_C}\bigr )}^{2(\zeta-\zeta_{loc})} \: 
       + \: b_3 { \bigl ( \frac{L_1}{L} \bigr )}^2
       {\bigl (\frac{L}{L_C}\bigr )}^{3(\zeta-\zeta_{loc})} \:
       + \: ... \: }} \label{eq:LSexp2}        
\end{eqnarray}
where $g_1(\alpha_o) = \partial g(\alpha_o)/ \partial \alpha$, 
$g_2(\alpha_o) = {\partial}^2 g(\alpha_o)/ {\partial \alpha}^2$, ..., and 
$b_2 = g(\alpha_o) g_2(\alpha_o)/(2 g_1(\alpha_o)^2)$, 
$b_3 = g(\alpha_o)^2 g_3(\alpha_o)/(6 {g_1(\alpha_o)}^3)$, ..., and, 
\begin{equation}
       \sigma_{M}  =  \sqrt{\frac{2\gamma E}{m g(\alpha_o) L_1}} 
       \label{eq:SM}\:,
        \qquad
       L_1 = c^{\ast} \: \frac{\beta}{m}  \: 
       \frac{g_1(\alpha_o)}{g(\alpha_o)}  \label{eq:SML1}   
\end{equation}
are all constants.  Equation (\ref{eq:LSexp2}) provides a large-size asymptotic series 
expansion of the size effect because the terms containing non  zero powers of $L$ in 
denominator vanish for $L \rightarrow \infty$. Note that Eq. (\ref{eq:LSexp2}) is expected 
to diverge for structure sizes $L \rightarrow 0$ as shown in Fig. \ref{fig:fig3}.  
A first-order asymptotic approximation at large sizes can be obtained by truncating the series 
after the linear term:
\begin{equation}
	\sigma_N =  \sigma_{M} \: 
       \sqrt{\frac{\biggl (1 \: + \: 
       \bigl (\frac{L}{L_C} \bigr )^{2(\zeta-\zeta_{loc})} \biggr )^{1/2}}
       {\frac{L}{L_1} \: + \: {\bigl (\frac{L}{L_C}\bigr )}^{\zeta-\zeta_{loc}} \: }} 
	\label{eq:LS1st}
\end{equation} 

The main consequence of the anomalous roughening in the case of large structure sizes is that 
the nominal strength is expected to decrease as 
$\sigma_N \sim L^{-1/2 +(\zeta-\zeta_{loc})/2}$ (as shown in Fig. \ref{fig:fig3}).  This result 
disagrees with the size effect proposed by Ba\v{z}ant \cite{Baz97c}  where the nominal strength 
of large structures decreases as $\sigma_N \sim L^{-1/2}$ which is the theoretical size effect 
of LEFM.  The difference originates in the fact that, for an anomalous roughening, the critical 
resistance to crack growth $G_{RC}$  [Eq. (\ref{eq:Grc})] is expected to evolve as a power law 
$G_{RC} \sim L^{\zeta-\zeta_{loc}}$ for large structure sizes while in LEFM, the critical 
resistance $G_{RC}$ is assumed to be constant (\ie~independant of the specimen size).  Hence, 
the size effect on the nominal strength obtained for an anomalous roughening is weaker than the 
size effect in LEFM.
% 
%----Small-size asymptotic expansion--------------------------------------------------------
%
\protect
\subsection{Small-size asymptotic expansion of the size effect}
\label{sec:Small_size}

In Ba\v{z}ant's theory \cite{Baz97c}, no size effect is expected for small structure sizes 
($L \rightarrow 0$); this is the domain of the strength theory.  A possible justification is 
that, in small structures, the fracture process zone fills the whole volume of the structure 
and hence, there is no stress concentration and, as a consequence, failure occurs with no 
crack propagation.

Such an argument can be also obtained from the link between anomalous roughening of crack 
surfaces and material fracture properties.  Indeed, in small structure sizes (\ie~$L \ll L_C$), 
the roughness being negligible, the virtual crack front length tends to its projected 
length $L$.  This implies that, for small structures, the actual crack surfaces produced during 
a crack advance $\Delta a_{sat}$ are not so different from the projected one: 
$A_r \simeq L\Delta a_{sat}$.  Hence, substituting $A_r \simeq L\Delta a_{sat}$ into the 
fracture criterion [Eq. (\ref{eq:Crit})] yields the crack length increment: 
$\Delta a_{sat} = n 2\gamma/G_d = c^{\ast}$ for $L \ll L_C$.  In other terms, the effective size 
of the process zone tends to the material length for small structure sizes.  Thus, when the 
material length $c^{\ast}=d-a_o$ (Fig. \ref{fig:fig2}), the fracture process zone occupies the 
entire ligament of the structure.  

On the basis of Ba\v{z}ant's theory \cite{Baz97c} and in order to obtain a small-size asymptotic 
expansion of the size effect, let us now introduce a new variable and a new function:
\begin{equation}
	\eta = \frac{1}{\theta} = \frac{d}{\Delta a_{sat}} \:, \qquad
	\varphi(\alpha_o,\eta) =  \frac{g(\alpha_o+\theta)}{\theta} = 
	\eta \: g(\alpha_o+1/\eta)
	\label{eq:Phi}
\end{equation} 
The function $\varphi(\alpha_o,\eta)$ corresponds to the dimensionless energy release rate 
function of the inverse relative size of the process zone.   Substituting Eq. (\ref{eq:Phi}) 
into Eq. (\ref{eq:Sn-d}) and expanding $\varphi(\alpha_o,\eta)$ into Taylor series around the 
point $(\alpha_o, 0)$ since $lim \: \eta = 0$ when $d$ or $L \rightarrow 0$, yields the nominal 
strength: 
\begin{eqnarray}
       \sigma_N & = & \sqrt{\frac{E\:G_{RC}}{c^{\ast}}} \:
       \biggl [ \: \varphi(\alpha_o,0) + \varphi_1(\alpha_o,0) \: \eta + 
       \varphi_2(\alpha_o,0) \: \frac{\eta^2}{2!} +
       \varphi_3(\alpha_o,0) \: \frac{\eta^3}{3!} + ...\: \biggr ]^{-1/2} 
       \label{eq:SSexp1}\\
       		& = & \sigma_{M^{\prime}} \: 
       \sqrt{\frac{\biggl (1 \: + \: 
       \bigl (\frac{L}{L_C} \bigr)^{2(\zeta-\zeta_{loc})} \biggr )^{1/2}}
       {1 \: + \: \frac{L}{L_2} \: + \: 
       \: c_2 {\bigl (\frac{L}{L_2}\bigr )}^2 \: 
       + \: c_3 { \bigl ( \frac{L}{L_2} \bigr )}^3 \: + \: ... \: }} 
       \label{eq:SSexp2}        
\end{eqnarray}
where $\varphi_1(\alpha_o,0) = \partial \varphi(\alpha_o,0)/ \partial \eta$, 
$\varphi_2(\alpha_o,0) = {\partial}^2 \varphi(\alpha_o,0)/ {\partial \eta}^2$, ..., 
and 
$c_2 = \varphi_2(\alpha_o,0) \varphi(\alpha_o,0)^2/(2 \varphi_1(\alpha_o,0)^2)$, 
$c_3 = \varphi_3(\alpha_o,0) \varphi(\alpha_o,0)^3/(6 {\varphi_1(\alpha_o,0)}^3)$, 
..., and, 
\begin{equation}
       \sigma_{M^{\prime}}  =  
       \sqrt{\frac{2\gamma E}{\varphi(\alpha_o,0) c^{\ast}}}\:,
        \qquad
       L_2 =  c^{\ast} \:  
       \frac{\varphi(\alpha_o,0)}{m \: \varphi_1(\alpha_o,0)}  
       \label{eq:SML2}   
\end{equation}
are all constants.  Equation (\ref{eq:SSexp2}) provides a small-size asymptotic series 
expansion of the size effect and is plotted in Fig. \ref{fig:fig3}.  When $L \rightarrow 0$, 
the nominal strength tends to a constant (\ie~$\sigma_{M^{\prime}}$ as expected in the case of 
a strength theory \cite{Baz97c}) but diverges from the asymptotic behavior of the size effect 
obtained in the case of large structure sizes [Eq. (\ref{eq:LSexp2})].  
%
%----Approximate size effect--------------------------------------------------------
%
\protect
\subsection{Approximate size effect}
\label{sec:Approx}

Now, the main problem consists in interpolating between the large-size [Eq. (\ref{eq:LSexp2})] 
and the small-size [Eq. (\ref{eq:SSexp2})] asymptotic series expansion in order to obtain an 
approximate size effect valid everywhere.  The theory of intermediate asymptotics \cite{Ben99} 
is not easily applicable in our case.  Nevertheless, it is interesting to observe in 
Fig. \ref{fig:fig3} that a satisfactory approximate size effect can be obtained by truncating 
the small-size asymptotic series expansion [Eq. (\ref{eq:SSexp2})] after the linear term:
\begin{equation}
	\sigma_N =  \sigma_{max} \: 
       \sqrt{\frac{\biggl (1 \: + \: 
       \bigl (\frac{L}{L_C} \bigr )^{2(\zeta-\zeta_{loc})} \biggr )^{1/2}}
       {1 \: + \: \frac{L}{L_o} \: }} 
	\label{eq:SEapp}
\end{equation} 
where the contants $\sigma_{max}$ and $L_o$ will be discussed in what follows.  Indeed, 
Eq. (\ref{eq:SEapp}) allows to obtain the transition between a horizontal asymptote 
characterizing the strength theory for which there is no size effect, and a decreasing 
asymptote, corresponding to a power law of exponent: 
$-1/2+(\zeta-\zeta_{loc})/2$ for large structure sizes.  A possible justification is that, 
in the large-size first-order expansion [Eq. (\ref{eq:LS1st})], the second term in numerator 
has no influence on the power law $\sigma_N \sim L^{-1/2 +(\zeta-\zeta_{loc})/2}$ for large 
sizes but induces a divergence at small sizes.

One limitation of the approximate size effect is that in Eq. (\ref{eq:SEapp}), the values of 
$\sigma_{M}$ and $\sigma_{M^{\prime}}$, and the values of $L_1$ and $L_2$, are assumed to be 
equal and respectively characterized by $\sigma_{max}$ and $L_o$. Indeed, the estimate of these 
values would be surely different from large size data or from small size data. Thus, the 
approximate size effect only gives the shape of the size effect relation on the nominal 
strength but does not allow for a determination of the parameter values $\sigma_{max}$ and 
$L_o$.  Only the crossover length $L_C$ and the scaling exponents $\zeta_{loc}$ and $\zeta$ are 
univocally determined from the roughness analysis.  However, it seems reasonnable to assume 
that the difference between $\sigma_{M}$ and $\sigma_{M^{\prime}}$ will be masked by the scatter 
of the experimental data.  On the other hand, it seems difficult to compare the crossover 
lengths $L_1$ and $L_2$ (characterized by $L_o$ in Eq. (\ref{eq:SEapp})) to the crossover length 
$L_C$ using their analytical expressions, because the former is deduced from a mechanical 
approach and the latter from a roughness analysis.  Nevertheless, from a physical point of view, 
both have the same meaning.  For small structure sizes, \ie~$L \ll L_C$ or $L_o$, the energy 
released by the structure is negligible, while for large sizes, \ie~$L \gg L_C$ or $L_o$, the 
energy released is dominant.  Hence, it seems reasonnable to assume that both crossover lengths 
are of the same order of magnitude (it is the case in Fig. \ref{fig:fig3} where it is assumed 
that $L_o=L_C$).
%
%%%%%%%%%This is new%%%%%%%%%%%%%%%%%%%%%%%%%%%%%%%%%%%%%%%%%%%%%%%%%%%  
%
%----Experiment--------------------------------------------------------  
%
\protect
\section{Experiment}
\label{sec:Exp}
In the following the experimental setup of mode I fracture tests on a quasibrittle material, 
wood, is described.  Two wood species have been tested: Maritime pine 
(\textit{Pinus pinaster Ait}) and Norway spruce (\textit{Picea Abies L.}).  The average oven 
dry specific densities were respectively $0.55$ and $0.40$ and the moisture content of
all specimens was measured between 11 and 13 \%.  Tests were made on modified tapered double 
cantilever beam specimens (TDCB) oriented along the  longitudinal-tangential directions of 
wood (Fig. \ref{fig:fig2}).  Six sets of geometrically similar specimens characterized by 
their width $L =7.5, 11.3, 15.0, 22.5, 30.0$ and $60.0 \:mm$ have been used.  A straight notch 
is machined with a band saw (thickness 2 $mm$) and prolonged a few millimeters with a razor 
blade (thickness 0.2 $mm$).  Fracture is obtained through uniaxial tension at a constant 
opening rate.  The fracture surfaces are generated along the average longitudinal-radial plane 
of wood.  Load-deflection values were continuously recorded during the tests.  

The roughness analysis \cite{Mor98}, performed from these fracture tests, has shown an 
anomalous roughening [Eq. (\ref{eq:Anomalous})] of the crack surfaces driven by the global 
and local roughness exponents, $\zeta=1.35\pm 0.10$ and $\zeta_{loc}=0.88\pm 0.07$ in the 
case of pine, and, $\zeta=1.60\pm 0.10$ and $\zeta_{loc}=0.87\pm 0.07$ in the case of spruce.  
Hence, the experimental \textit{R}-curve behaviors and the size effects on critical energy 
release rates and nominal strengths should be described respectively by Eq. (\ref{eq:Gr}), 
Eq. (\ref{eq:Grc}) and Eq. (\ref{eq:SEapp}). 

From an equivalent linear elastic approach (which is the frame of the model described in 
Sec. \ref{sec:Scaling} and Sec.\ref{sec:Sizeffect}) where the crack lengths can be estimated 
from the unloading compliance of the specimens , the elastic energy release rate $G_R$ 
are computed from the load-deflection curves for any crack length increment $\Delta a$.  
Two examples of the energy release rate evolution $G_R$ as a function of the crack length 
extension $\Delta a$ obtained for both wood species and for a specimen size $L=11.3\:mm$ are 
given in Fig. \ref{fig:fig4}.  An \textit{R}-curve behavior is observed for both wood species, 
\ie~a pronounced evolution of the resistance to crack growth as a function of the equivalent 
crack length increment \cite{Law93}.  After a characteristic propagation distance 
($\Delta a_{sat}\simeq 23\:mm$ for pine (a) and $40\:mm$ for spruce (b)) which corresponds 
approximately to the ultimate load $P_u$, the energy release rate becomes independent of the 
crack length increment $\Delta a$.  The post \textit{R}-curve behavior arises at constant 
resistance to crack growth and corresponds to critical resistance $G_{RC}$.   
In Figure \ref{fig:fig4} the \textit{R}-curves and post \textit{R}-curve behaviors are fitted 
with the Eq. (\ref{eq:Gr}).  These fits are obtained in keeping three free parameters: 
the specific surface energy $\gamma$, the ratio $A/{l_o}^{1-\zeta_{loc}}$ and the scaling 
exponent $(\zeta-\zeta_{loc})/z$.  The expected \textit{R}-curve behavior [Eq. (\ref{eq:Gr})] 
provides a good description of the increase of the experimental resistances to crack growth.  
For both wood species, the fitted exponents $(\zeta-{\zeta}_{loc})/z$ are close to those 
measured from the roughness analysis \cite{Mor98} which are given in brackets in  
Fig. \ref{fig:fig4}.  Note that the \textit{R}-curve is more pronouced in pine than in spruce 
(\ie~the critical energy release rate $G_{RC}$ is greater for a smaller crack length increment 
$\Delta a_{sat}$ than in spruce).  This trend has also been observed for the other specimen 
sizes $L$.    

On the other hand, for both wood species, it has been shown that the sizes $L$ of tested 
specimens were greater than the crossover lengths $L_C$ \cite{Mor00} [defined in 
Eq.(\ref{eq:Grc}) and Eq. (\ref{eq:SEapp}) relative to the size effect].  
Thus, according to Eq. (\ref{eq:Grc}) and for $L \gg L_C$, the size 
effect on the critical energy release rates of both wood species is expected to evolve as 
a power law: $G_{RC} \sim L^{\zeta-\zeta_{loc}}$.  The critical energy release rates obtained 
for both wood species are plotted in Fig. \ref{fig:fig5} versus the characteristic specimen 
sizes $L$.  Experimental size effects are fitted by a power law $G_{RC}(L) \sim L^\beta$ 
whose the exponents $\beta=0.42$ for pine (a) and $\beta=0.64$ for spruce (b) are in fair 
agreement with the expected exponents $\zeta-\zeta_{loc}=0.47 \pm 0.17$ for pine (a) 
and $0.73 \pm 0.17$ for spruce (b).  Note that the size effect on critical energies of spruce 
is stronger than the one obtained for pine.  
   
In the same way, in Fig. \ref{fig:fig6}, the nominal strengths $\sigma_N$ obtained for both 
wood species are plotted versus the characteristic specimen sizes $L$.  According to 
Eq. (\ref{eq:SEapp}) and, as previously mentioned for $L \gg L_C$, the size effect on nominal 
strengths is expected to evolve as: $\sigma_N \sim L^{-1/2 +(\zeta-\zeta_{loc})/2}$.  
Note that latter asymptotic behaviour is obtained with the assumption $L_o=L_C$ in 
Eq. (\ref{eq:SEapp}) as previously discussed in Sec \ref{sec:Approx}.  
As shown in Fig. \ref{fig:fig6}, simple power law fits of the experimental nominal strengths 
give exponent which are in good agreement with those expected, 
\ie~$-\frac{1}{2}+\frac{\zeta-\zeta_{loc}}{2}=-0.27\pm0.09$ for pine (a) and $-0.14\pm0.09$ 
for spruce (b). 
By comparison of the size effects on the critical energy release rates and on the nominal 
strengths it is appeared that, as expected intuitively, the more the size effect on critical 
energies is important the more the size effect on nominal strengths is weak.  
%
%----Conclusion--------------------------------------------------------  
%
\protect
\section{Conclusion}
\label{sec:Conclu}

From an energy-based analysis, a link between the scaling laws describing the fracture surfaces 
and the size effect on the nominal strength of the structures is proposed.  On the basis of a 
Family-Vicsek scaling, which has been shown to reflect a purely elastic brittle fracture behavior 
\cite{Mor00}, the size effect obtained is in agreement with the classical size effect of linear 
elastic fracture mechanics: $\sigma_N \sim L^{-1/2}$ [Eq. (\ref{eq:SizeFam})].  In the case of 
an anomalous scaling, reflecting the fracture behavior of quasibrittle materials \cite{Mor00}, 
an asymptotic analysis allows to estimate the size effect relation on the nominal strength, and 
especially, the crack length increment for which the maximum load is reached.  From large-size 
and small-size asymptotic series expansion of the size effect, an approximate size effect 
[Eq. (\ref{eq:SEapp})] of general validity is proposed. This relation represents a smooth 
transition from the case of no size effect, for small structure sizes, to a power law size 
effect which appears weaker than the size effect of LEFM. Thus, in the case of quasibrittle 
materials, the approximate size effect relation [Eq. (\ref{eq:SEapp})] is different from the 
\textit{classical} size effect law proposed by Ba\v{z}ant \cite{Baz97c} and especially for large 
structure sizes where an asymptotic behavior $\sigma_N \sim L^{-1/2 +(\zeta-\zeta_{loc})/2}$ is 
predicted instead of the size effect of LEFM suggested by Ba\v{z}ant's theory.  The difference 
can be explained by the fact that, for an anomalous scaling of fracture surfaces, the critical 
resistance to crack growth at the maximum load evolves as a power law 
$G_{RC} \sim L^{\zeta-\zeta_{loc}}$ for large structure sizes while this resistance is assumed 
to be constant in LEFM.  Experiments performed on geometrically similar wood specimens of 
various sizes, for which anomalous roughening of crack surfaces has been observed previously 
\cite{Mor98}, show that the size effects on the critical resistances and on the nominal 
strengths are in fair agreement with the predicted asymptotic behaviors 
$G_{RC} \sim L^{\zeta-\zeta_{loc}}$ and $\sigma_N \sim L^{-1/2 +(\zeta-\zeta_{loc})/2}$. 
On the other hand, if one considers a \textit{weakly} anomalous roughening of the fracture 
surfaces (\ie~$\zeta \rightarrow \zeta_{loc}$), \textit{R}-curve [Eq. (\ref{eq:Gr})] and size 
effect on the critical resistance [Eq. (\ref{eq:Grc})] vanish progressively and as a 
consequence, the size effect characteristic of a quasibrittle material [Eq. (\ref{eq:SEapp})] 
tends to the classical size effect of a purely elastic brittle material 
[Eq. (\ref{eq:SizeFam})].  
Experiments on different kinds of quasi-brittle materials are currently being performed to test 
our predictions.  
%
%--References-------------------------------------------------------
%

%\end{multicols}
%
%--Captions-------------------------------------------------------
%
\newpage
\begin{figure}
\protect
\caption{Size effect on the critical resistance to crack growth $G_{RC}$ obtained 
	in the case of an anomalous scaling ($\zeta=1.3$, ${\zeta}_{loc}=0.8$, 
	$A=0.1$ and $l_o=1$: arbitrary values}
\label{fig:fig1}
\end{figure}
\begin{figure}
\protect
\caption{Geometrically similar Tapered Double Cantilever Beam (TDCB) fracture 
	specimens of different sizes  
	  $L=7.5,11.3,15.0,22.5,30.0$ and $60.0$ mm.}
\label{fig:fig2}
\end{figure}
\begin{figure}
\protect
\caption{Approximate size effect on the nominal strength (solid curve, 
	Eq. (\ref{eq:SEapp})) and asymptotic series expansions (dashed curves, 
	Eq. (\ref{eq:LSexp2}) and Eq. (\ref{eq:SSexp2}))}
\label{fig:fig3}
\end{figure}
\begin{figure}
\protect
\caption{Examples of \textit{R}-curves $G_{R}(\Delta a)$ respectively obtained for 
	pine (a) and spruce (b) from specimens of characteristic size $L=11,3\:mm$.}
\label{fig:fig4}
\end{figure}
\begin{figure}
\protect
\caption{Size effect on critical energy release rate $G_{RC}$ respectively for 
	pine (a) and spruce (b). 
	The expected slopes from the roughness analysis \cite{Mor98} are 
	$\zeta-\zeta_{loc}=0.47 \pm 0.17$ for pine (a) 
	and $0.73 \pm 0.17$ for spruce (b).}
\label{fig:fig5}
\end{figure}
\begin{figure}
\protect
\caption{Size effect on nominal strength respectively for pine (a) and spruce (b). 
	The expected slopes from the roughness analysis \cite{Mor98} are 
	$-\frac{1}{2}+\frac{\zeta-\zeta_{loc}}{2}=-0.27\pm0.09$ for pine (a) 
	and $-0.14\pm0.09$ for spruce (b).}
\label{fig:fig6}
\end{figure}
\newpage
\psfig{figure=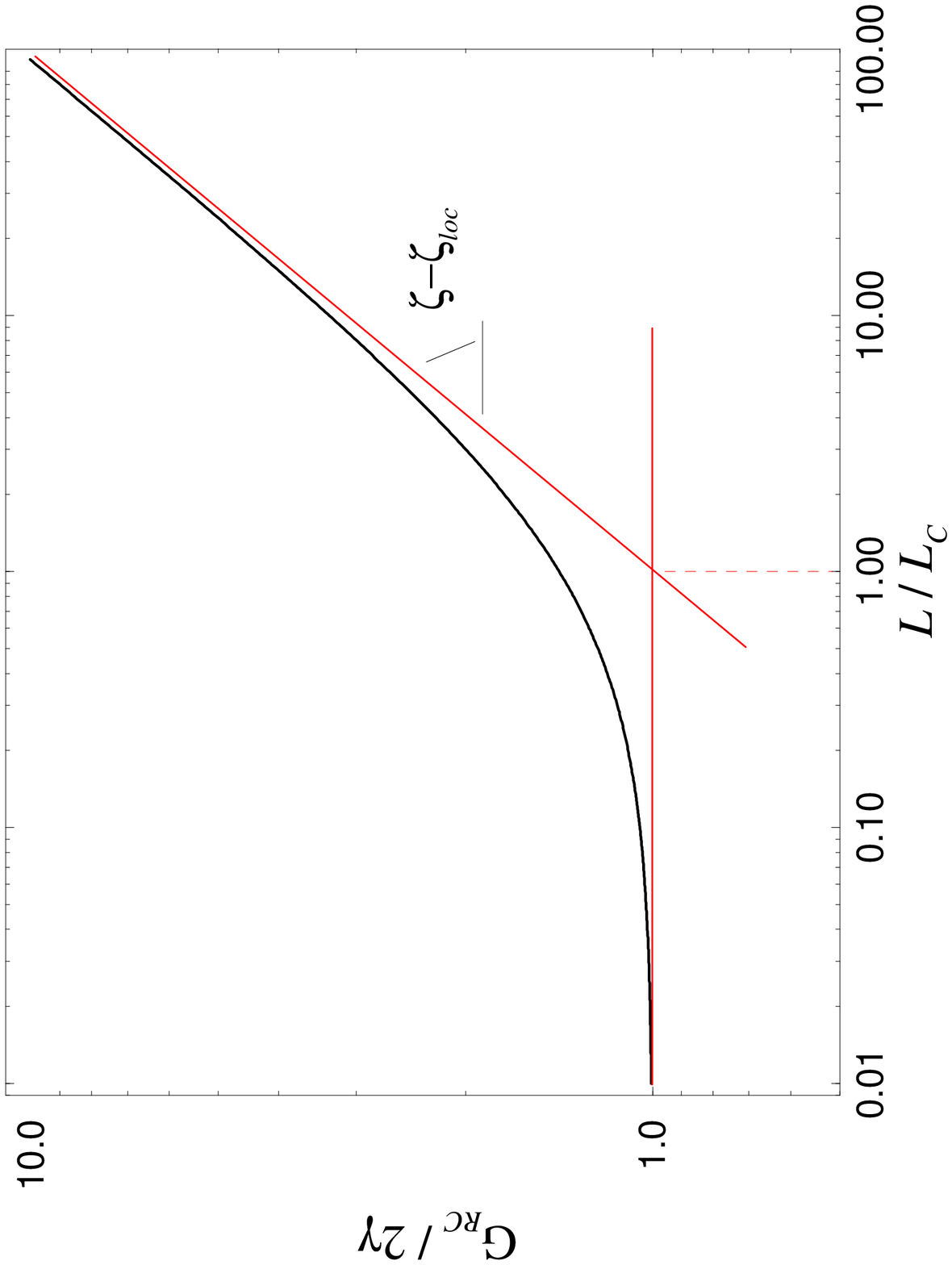,width=15cm}
\vskip 2cm
Figure1-Morel et al.-PRB
\newpage
\psfig{figure=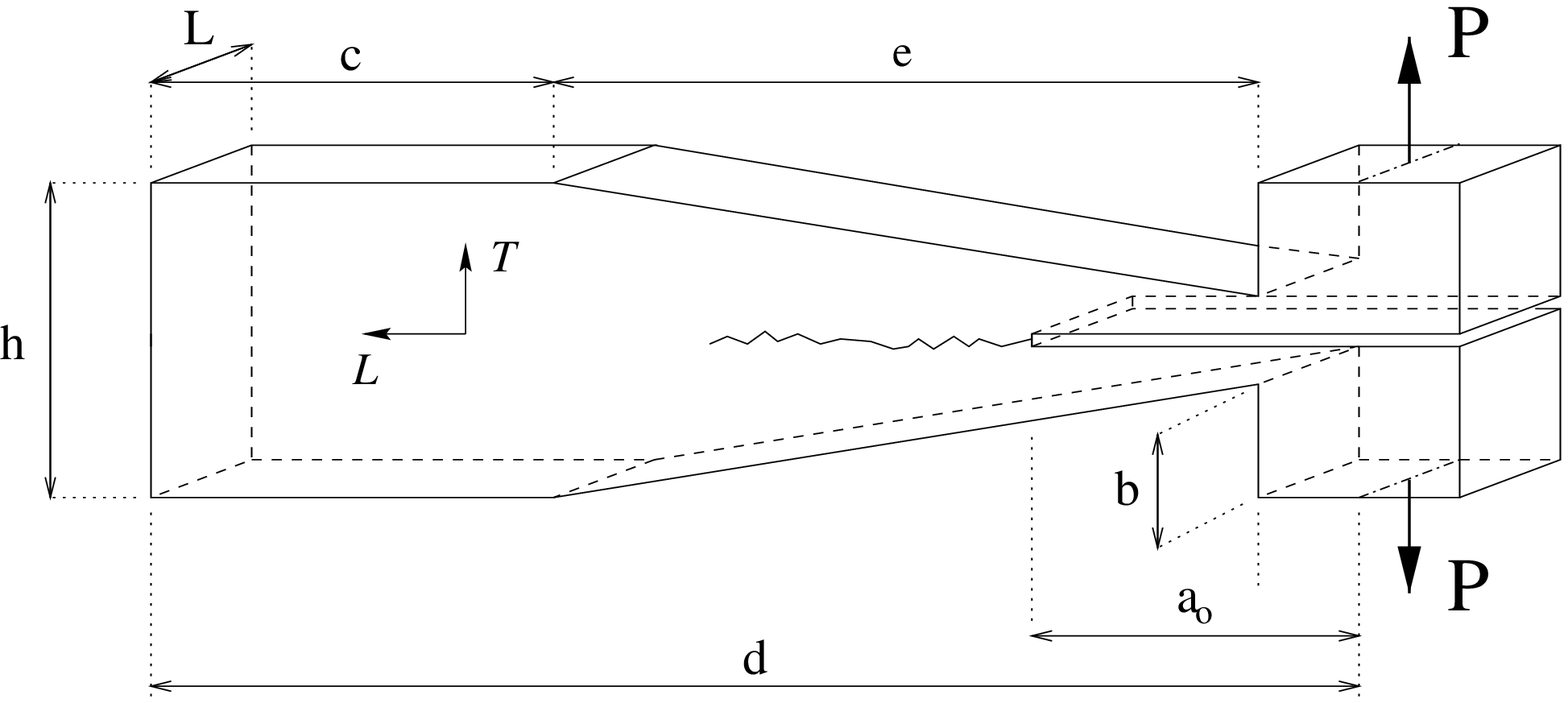,angle=90,width=8cm}
\vskip 2cm
Figure2-Morel et al.-PRB
\newpage
\psfig{figure=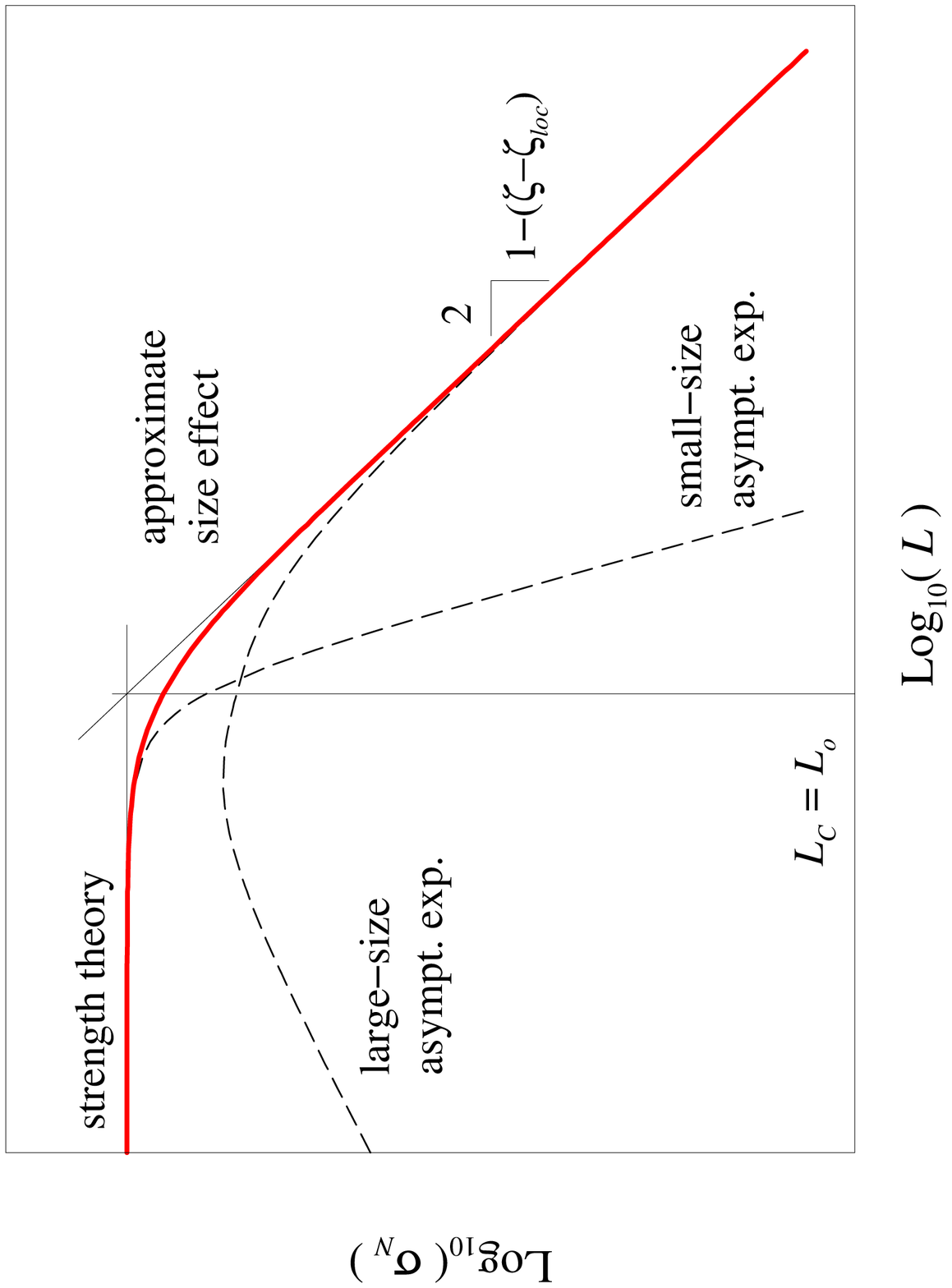,width=15cm}
\vskip 2cm
Figure3-Morel et al.-PRB
\newpage
\psfig{figure=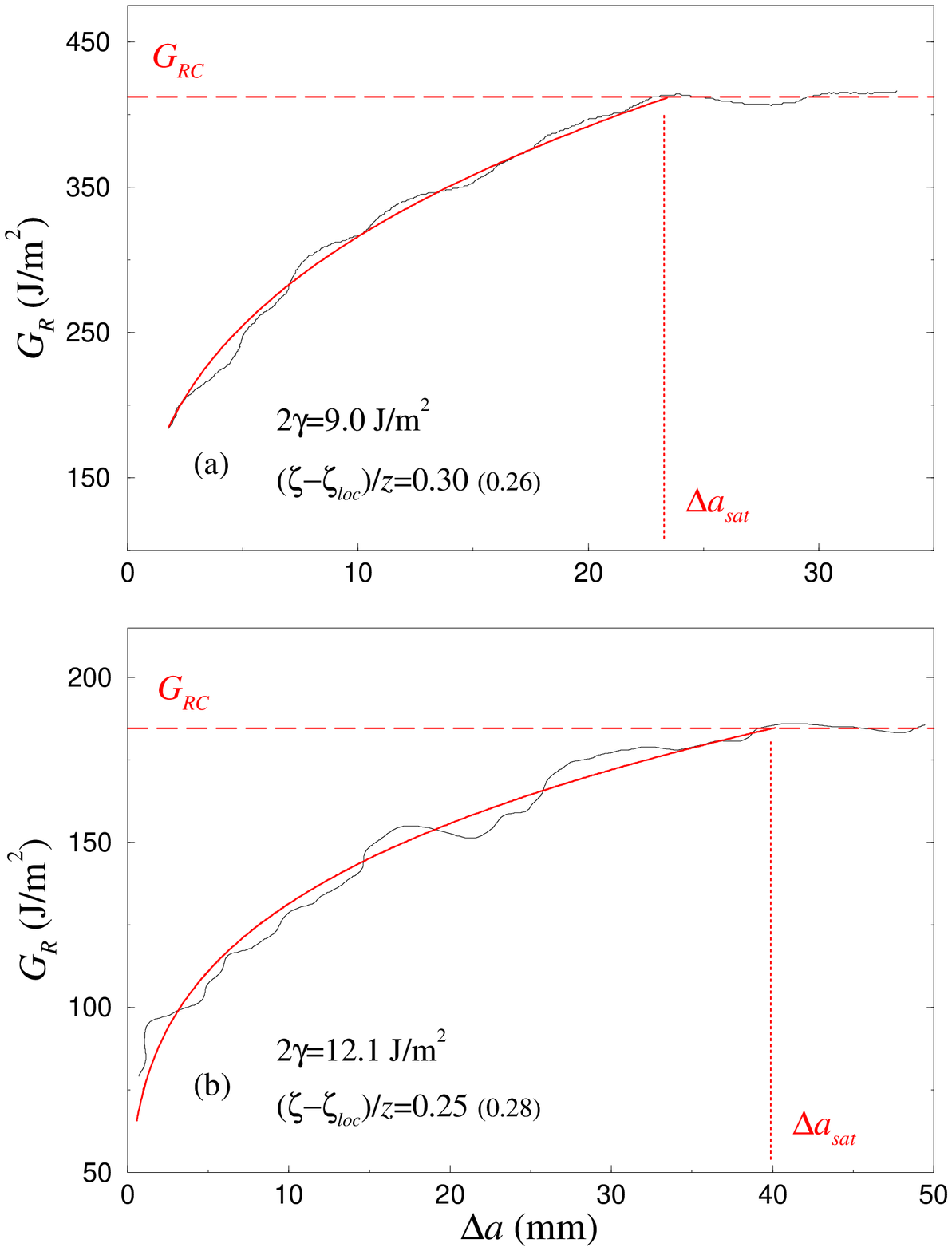,width=15cm}
\vskip 2cm
Figure4-Morel et al.-PRB
\newpage
\psfig{figure=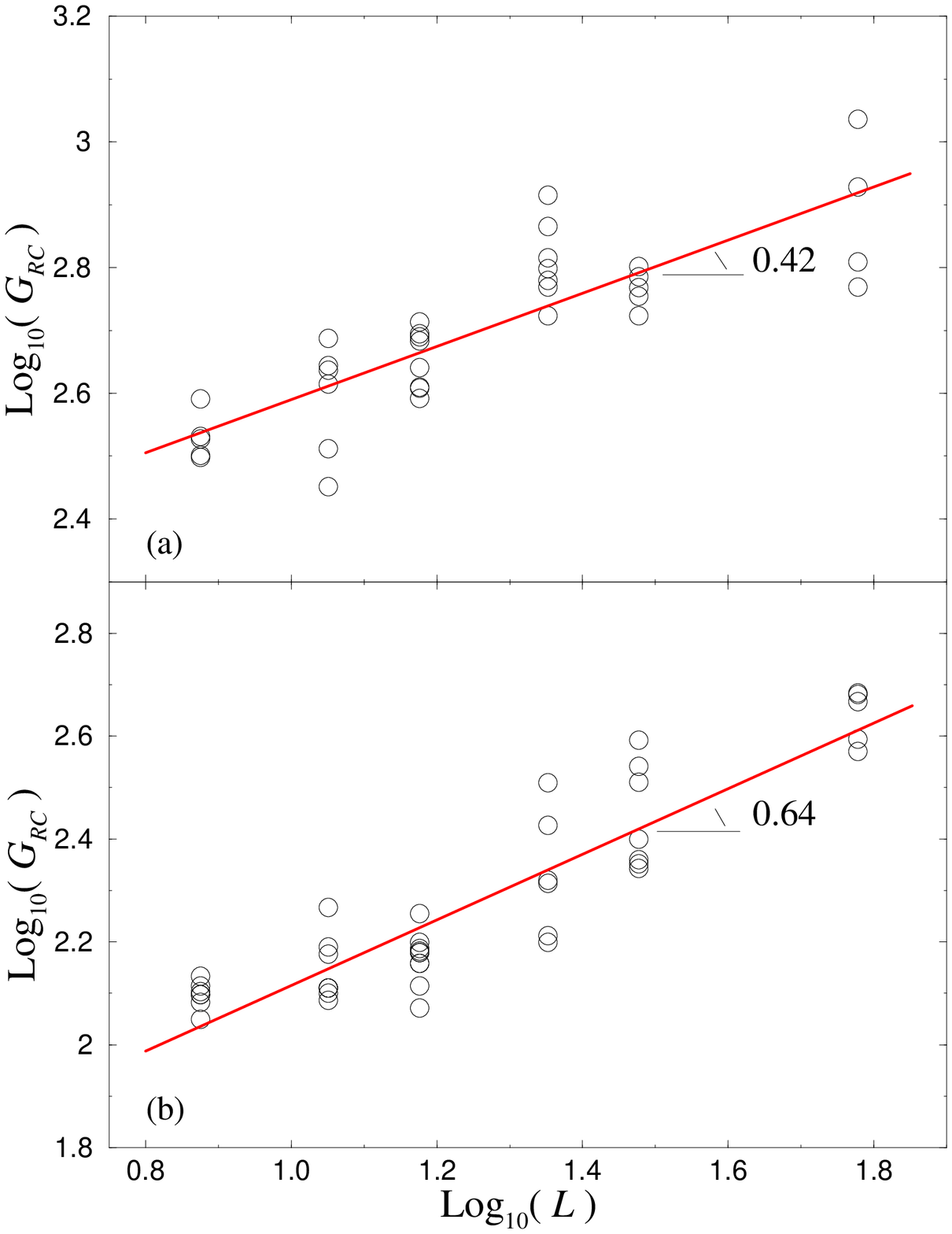,width=15cm}
\vskip 2cm
Figure5-Morel et al.-PRB
\newpage
\psfig{figure=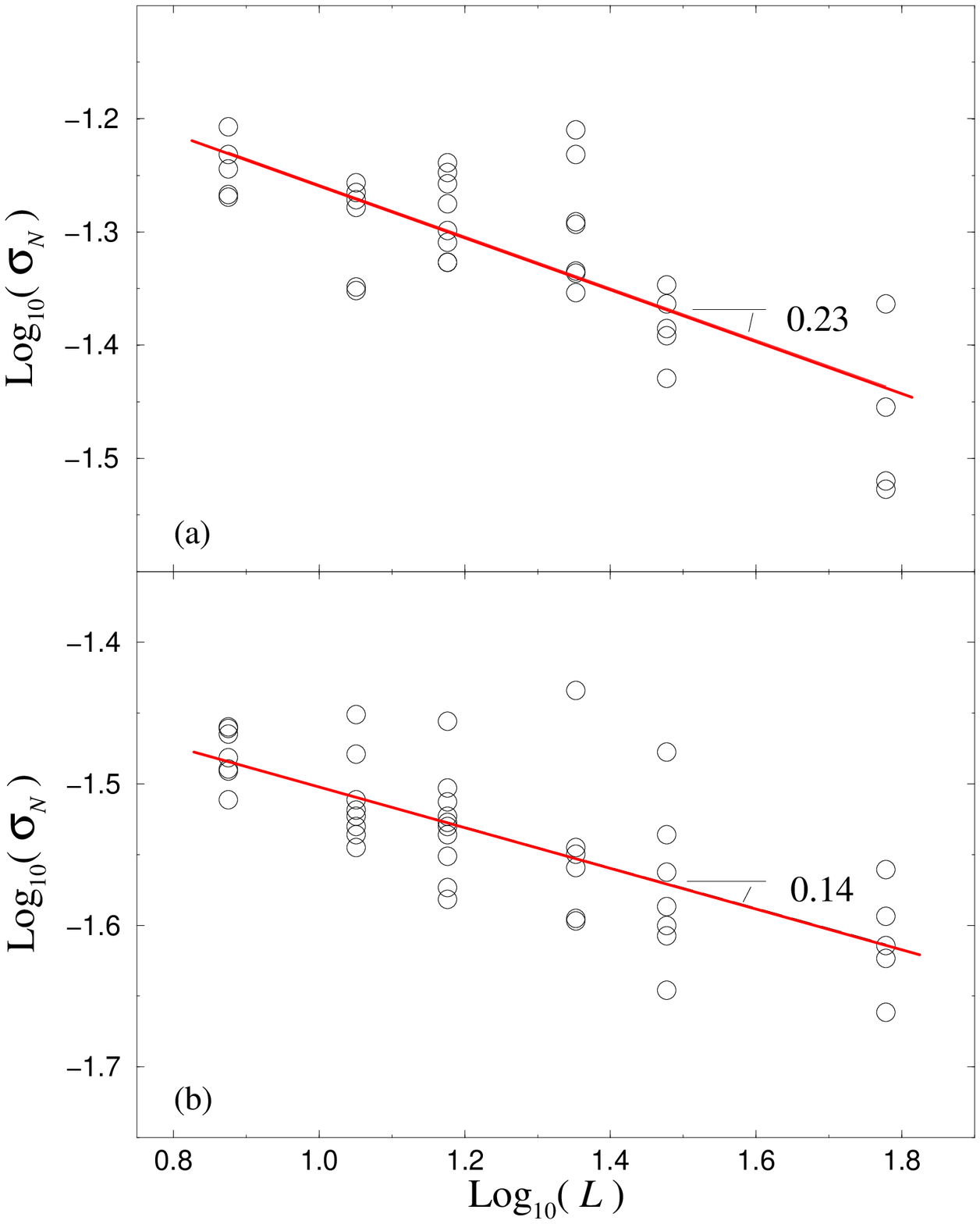,width=15cm}
\vskip 2cm
Figure6-Morel et al.-PRB
\end{document}